\documentclass[twocolumn,superscriptaddress,showpacs,pra]{revtex4}
\usepackage{epsfig}
\usepackage{epstopdf}
\usepackage{delarray}
\usepackage{amsmath, amssymb}
\usepackage{bm}
\usepackage{graphicx}% Include figure files
\usepackage{placeins}
\usepackage{color}
\begin{document}
\title{Compressive Spectral Renormalization Method}

\author{Cihan A. Bay\i nd\i r}
\email{cihan.bayindir@isikun.edu.tr}
\affiliation{Engineering Faculty, Isik University, Istanbul, Turkey}

%\date{\today}
\begin{abstract}
In this paper a novel numerical scheme for finding the sparse self-localized states of a nonlinear system of equations with missing spectral data is introduced. As in the Petviashivili's and the spectral renormalization method, the governing equation is transformed into Fourier domain, but the iterations are performed for far fewer number of spectral components ($M$) than classical versions of the these methods with higher number of spectral components ($N$). After the converge criteria is achieved for $M$ components, $N$ component signal is reconstructed from $M$ components by using the $l_1$ minimization technique of the compressive sampling. This method can be named as compressive spectral renormalization (CSRM) method. The main advantage of the CSRM is that, it is capable of finding the sparse self-localized states of the evolution equation(s) with many spectral data missing. 

\end{abstract}
\maketitle

%%%%%%%%%%%%%%%%%%%%%%%%%%%%%%% main %%%%%%%%%%%%%%%%%%%%%%%%%%%%%
\section{\label{sec:level1} Introduction}
Missing spectral data in many branches of physics limits the applicability of analysis tools and devices and assessment techniques. These branches include but are not limited to optics, geophysics, electromagnetics and communication technology \cite{Stoica, Schoellhamer}. Although shorter wavelengths are more vulnerable to attenuation compared to longer wavelengths, longer wavelengths can also be lost during their propagation in medium therefore any spectral component of a signal may be lost \cite{Thiele}. Different approaches are proposed to deal with problems that occur due to missing spectral data. Few of these approaches can be summarized as below: A modification of the singular spectrum analysis that analyzes time series with missing data is proposed in \cite{Schoellhamer}. The missing data recovery via a nonparametric iterative adaptive approach and nonparametric spectral analysis with missing data via the expectation maximization algorithms are two other tools used in literature to overcome this problem \cite{Stoica}. The method of Marquardt is applied to a single $26$-parameter equation, which models known long-wavelength loss mechanisms for rapid and accurate modeling of the spectral-loss profiles of lightguide fibers, in \cite{Walker}. A compressive sensing (CS) based approach for stationary and non-stationary stochastic process power spectrum estimation with missing data is recently proposed in \cite{Comerford}. 

It remains an open question what the stable self-localized solutions would be in a nonlinear field with missing spectral data. To overcome this problem, we propose a novel numerical scheme that can attain the stable sparse self-localized solutions of the nonlinear system of equations with many missing spectral data. We test the applicability of the proposed method on a 1D Schr\"{o}dinger-like equation which is widely used as a model equation in optics, hydrodynamics, quantum mechanics, Bose-Einstein condensation \cite{Ablowitz, Zakharov1968, Zakharov1972}. The proposed method utilizes iterations in the Fourier domain for a far fewer number of spectral components ($M$) than classical versions of the Petviashivili's method or spectral renormalization method (SRM) with higher number of spectral components ($N$). After the converge criteria is achieved for $M$ components, the signal with $N$ components can be reconstructed from $M$ components by the $l_1$ minimization technique of the compressive sampling. The name proposed for this method is compressive spectral renormalization (CSRM) method. Compared to SRM, the main advantage of the CSRM is that, it is capable of finding the sparse self-localized states of the evolution equation(s) with far fewer spectral data. For example for fiber optical communications, where some data is lost during the propagation of the optical pulse or considering memory and time constraints they may be intentionally ignored, CSRM can be used to find self localized states of the system of equations studied. We discuss the implementation of the proposed method and its advantages and limitations using single and dual soliton solutions of the NLS and an NLS-like equation with a potential used to model the photorefractive lattice solitons and with saturable nonlinearity.

\section{\label{sec:level1}Methodology}

\subsection{Review of the Spectral Renormalization Method} 

Self-localized solutions of many nonlinear systems can be found by various computational techniques such as shooting, self-consistency and relaxation \cite{Ablowitz}. Another method known is the Petviashvili's method which is based on transforming the governing nonlinear equation into Fourier space, as in the case of general Fourier spectral schemes \cite{bay2009, Canuto, Karjadi2010,Karjadi2012, trefethen,  BayPRE1, BayPRE2, BayTWMS2016, Demiray2015, BayPLA, Bay_arxNoisyTun, Bay_arxNoisyTunKEE, Bay_arxEarlyDetectCS, Bay_arxChaotCurNLS}, and determining a convergence factor according to the degree of a single nonlinear term \cite{Petviashvili, Ablowitz}. This method was introduced by Petviashvili and applied to Kadomtsev-Petviashvili (2D Korteweg de-Vries) equation \cite{Petviashvili}. Later, it has been developed and applied to many other systems which model many different phenomena such as lattice vortices, dark and gray solitons \cite{Ablowitz, Yang}. Since Petviashvili's method works well for nonlinearities with fixed homogeneity only, a novel spectral renormalization method (SRM) is proposed in \cite{Ablowitz, Fibich} which is capable of finding the localized solutions in waveguides with different types of nonlinearities. The SRM essentially transforms the governing equation into Fourier space and couples it to a nonlinear integral equation which is basically an energy conservation principle for the iterations in the Fourier space \cite{Ablowitz}. This coupling makes the initial conditions to converge to the self-localized solutions of the nonlinear system studied \cite{Ablowitz}. SRM is spectrally efficient, it can be applied to many different physical problems with different higher-order nonlinearities and is easy to implement \cite{Ablowitz}. Following \cite{Ablowitz}, we give a brief review of the SRM considering a 1D NLS-like equation as
\begin{equation}
i\zeta_z +  \zeta_{xx} -V(x)\zeta+ N(\left| \zeta \right|^2) \zeta =0
\label{eq01}
\end{equation}
where $z$ is the propagation direction of optical pulse, $x$ is the transverse coordinate, $i$ denotes the imaginary number and $\zeta$ is complex amplitude of the optical field \cite{Ablowitz, Agarwal}. Using the ansatz, $\zeta(x,z)=\eta(x,\mu) \textnormal{exp}(i\mu z)$, where $\mu$ shows the soliton eigenvalue, the NLS-like equation becomes
\begin{equation}
-\mu \eta +  \eta_{xx} -V(x)\eta+ N(\left| \eta \right|^2) \eta =0
\label{eq02}
\end{equation}
Furthermore the 1D Fourier transform of $\eta$ can be obtained by
\begin{equation}
\widehat{\eta} (k)=F[\eta(x)] = \int_{-\infty}^{+\infty} \eta(x) \exp[i(kx)]dx
\label{eq03}
\end{equation}
For a zero optical potential, $V=0$, the 1D Fourier transform of Eq.~(\ref{eq02}) yields
\begin{equation}
\widehat{\eta} (k)=\frac{F \left[ N( \left| \eta \right|^2\eta) \right]}{\mu+\left| k \right|^2}
\label{eq04}
\end{equation}
This formula may be applied iteratively to find the self-localized solutions of the system studied, as first proposed by Petviashvili in \cite{Petviashvili}. However iterations of Eq.~(\ref{eq03}) may grow unboundedly or may tend to zero \cite{Ablowitz}. As proposed in \cite{Ablowitz}, this problem can be solved by introducing a new variable in the form $\eta(x)=\alpha \xi(x)$ which has a 1D Fourier transform $\widehat{\eta}(k)=\alpha \widehat{\xi}(k)$. With these substitutions, Eq.~(\ref{eq04}) becomes
\begin{equation}
\widehat{\xi} (k)=\frac{F\left[ N( \left| \alpha \right|^2 \left| \xi \right|^2 ) \xi \right]}{\mu+\left| k \right|^2}=R_{\alpha}[\widehat{\xi} (k)]
\label{eq05}
\end{equation}
and thus the iteration scheme is given as
\begin{equation}
\widehat{\xi}_{j+1} (k)=\frac{F\left[N( \left| \alpha_j \right|^2 \left| \xi_j \right|^2 ) \xi_j \right]}{\mu+\left| k \right|^2}
\label{eq06}
\end{equation}
For the normalization part of the SRM, an algebraic condition on the parameter $\alpha $ can be obtained using the energy conservation principle. Multiplying both sides of Eq.~(\ref{eq05}) with the complex conjugate of $\widehat{\xi}(k)$, which is $\widehat{\xi}^*(k)$, and integrating to evaluate the total energy, the algebraic condition becomes 
\begin{equation}
\int_{-\infty}^{+\infty} \left|\widehat{\xi} (k)\right|^2 dk= \int_{-\infty}^{+\infty} \widehat{\xi}^* (k) R_{\alpha}[\widehat{\xi} (k)]dk  
\label{eq07}
\end{equation}
which is the normalization constraint that ensures the scheme to converge to a stable self-localized state, solitons. The procedure of obtaining self-localized solutions of a nonlinear system by the coupled equation analysis reviewed and summarized above is known as the spectral renormalization method (SRM) \cite{Ablowitz}. Starting with an initial condition in the form of a single or multi-Gaussians, Eq.~(\ref{eq04}) is applied to find the profile for next iteration step, then the normalization constraint given by Eq.~(\ref{eq07}) is applied. Iterations can be continued until the convergence of ${\alpha}$ is achieved.

Nonzero potentials ($V\neq0$) are widely accepted as models for various optical media such as nondefected or defected photonic crystals. Adding and substracting a $p \eta$ term with $p>0$ from Eq.~(\ref{eq02}) in order to avoid singularity of the scheme \cite{Ablowitz}, the 1D Fourier transform of Eq.~(\ref{eq02}) becomes
\begin{equation}
\widehat{\eta} (k)=\frac{(p+| \mu|)\widehat{\eta}}{p+\left| k \right|^2} -\frac{F[V \eta]-F \left[ N(\left| \eta \right|^2) \eta \right]}{p+\left| k \right|^2}
\label{eq08}
\end{equation}
which is the iteration scheme for a nonzero optical potential \cite{Ablowitz}. In this paper we are specifically interested in photorefractive solitons of practical use. Therefore, considering the 1D versions of the photorefractive solitons first reported in \cite{Segev}, we set the optical potential as $V=I_o \cos^2(x)$ and the nonlinear term as $N(\left| \eta \right|^2)=-1/(1+\left| \eta \right|^2)$. As before, one can define a new parameter $\eta(x)=\alpha \xi(x)$ and evaluate its Fourier transform as $\widehat{\eta}(k)=\alpha \widehat{\xi}(k)$. With these substitutions iteration formula reads
\begin{equation}
\begin{split}
\widehat{\xi}_{j+1} (k) &=\frac{(p+| \mu|)}{p+\left| k \right|^2}\widehat{\xi_j}-\frac{F[V \xi_j]}{p+\left| k \right|^2} \\
& +\frac{1}{p+\left| k \right|^2}  F\left[ \frac{\xi_j}{1+\left| \alpha_j \right|^2 \left| \xi_j \right|^2}\right] =R_{\alpha_j}[\widehat{\xi}_j (k)]
\label{eq09}
\end{split}
\end{equation}
The algebraic condition of the SRM for nonzero potential case can be attained by multiplying both sides of Eq.~(\ref{eq09}) with the complex conjugate of $\widehat{\xi}(k)$, which is $\widehat{\xi}^*(k)$, and integrating to evaluate the total energy the normalization constraint becomes
\begin{equation}
\int_{-\infty}^{+\infty} \left|\widehat{\xi} (k)\right|^2 dk= \int_{-\infty}^{+\infty} \widehat{\xi}^* (k) R_{\alpha}[\widehat{\xi} (k)]dk  
\label{eq10}
\end{equation}
As in the case of zero potentials, an initial condition in the form of a single or multi-Gaussians converges to self-localized states of the model equation when Eq.~(\ref{eq09}) is applied to find the profile for next iteration step and then the normalization constraint given by Eq.~(\ref{eq10}) is applied. Iterations can be continued until the convergence of ${\alpha}$ with a specified upper bound is achieved. A detailed discussion and application of SRM to 2D and second-harmonic generation problems can be seen in \cite{Ablowitz}.

\subsection{Review of the Compressive Sampling}
%\begin{doublespace}
\noindent Compressive sampling (CS) is an efficient sampling technique which exploits the sparsity of the signal for its reconstruction by using far fewer samples than the requirements of the Shannon-Nyquist sampling theorem  \cite{candes2006compressive, Candes2006}. Since its introduction to the scientific community, CS has been intensively studied as a mathematical tool in applied mathematics and physics and currently sampling in various engineering devices such as the single pixel video cameras and efficient A-D converters is performed using CS. We give a very brief summary of the CS in this section and refer the reader to \cite{candes2006compressive, Candes2006} for a comprehensive analysis.

Let $\zeta$ be a $K$-sparse signal with $N$ elements. This means that only $K$ of the $N$ elements of $\zeta$ are nonzero. Using orthonormal basis transformations with transformation a matrix ${\bf \Psi}$, $\zeta$ can be represented in transformed domain in terms of the basis functions. Typical orthogonal transformations used in the literature include but are not limited to Fourier, wavelet or discrete cosine transforms (DCT). Using the orthogonal transformation it is possible to express the signal as $\zeta= {\bf \Psi} \widehat{ \zeta}$ where $\widehat{ \zeta}$ is the coefficient vector. Discarding the zero coefficients and keeping the non-zero coefficients of $\zeta$, one can obtain $\zeta_s= {\bf \Psi}\widehat{ \zeta}_s$  where $\zeta_s$ denotes the signal with non-zero entries.

CS algorithm states that a $K$-sparse signal $\zeta$ which has $N$ elements can exactly be reconstructed from $M \geq C \mu^2 ({\bf \Phi},{\bf \Psi}) K \textnormal{ log (N)}$ measurements with a very high probability. In here $C$ is a positive constant and $\mu^2 (\Phi,\Psi)$ is the mutual coherence between the sensing ${\bf \Phi}$ and transform bases ${\bf \Psi}$ \cite{candes2006compressive, Candes2006}. Taking $M$ projections randomly and using the sensing matrix ${\bf \Phi}$ the sampled signal can be written as $g={\bf \Phi} \zeta$. Therefore the CS problem can be rewritten as
\begin{equation}
\textnormal{ min} \left\| \widehat{ \zeta} \right\|_{l_1}   \ \ \ \  \textnormal{under constraint}  \ \ \ \ g={\bf \Phi} {\bf \Psi} \widehat{ \zeta}
\label{eq14}
\end{equation}
where $\left\| \widehat{ \zeta} \right\|_{l_1}=\sum_i \left| \widehat{ \zeta}_i\right|$. Therefore among all signals that satisfy the given constraints above, the ${l_1}$ minimization solution of the CS problem is $\zeta_{{}_{CS}} ={\bf \Psi} \widehat{ \zeta}$.  $l_1$ minimization is only one of the tools that can be used for finding the solution of this optimization problem. The sparse signals can also be recovered using other optimization techniques such as the re-weighted $l_1 $ minimization or greedy pursuit algorithms \cite{candes2006compressive, Candes2006}. Details of the CS can be seen in \cite{candes2006compressive, Candes2006}.  

%\end{doublespace}
\subsection{Proposed Compressive Spectral Renormalization Method}
%\begin{doublespace}
%In this study a methodology which can significantly reduce the computational effort required for the spectral simulations of the sparse water waves is offered. 

\noindent In a classical SRM let $N$ be the number of the spectral components used to represent a self-localized solution of an evolution equation(s). First we select $M$ spectral components at random, where $M << N$, and apply SRM for those $M$ components. The random selection of the number $M$ needs to be done carefully depending on width of the $K$-sparse self-localized state since $M$ needs to satisfy the $M=O(K \log(N/K))$ condition of the CS. Starting from the initial conditions, iterations are performed for obtaining the convergent self-localized states for $M$ components. After the converge criteria is achieved for $M$ components, $N$ component signal is reconstructed from $M$ components by using the $l_1$ minimization technique of the CS. This method can be named as compressive spectral renormalization (CSRM) method. The advantage of the CSRM is that, it is capable of finding the sparse self-localized states of the evolution equation(s) with $N-M$ spectral data missing. In practice, for example in a typical photonic crystal, only few of the spectral data would be expected to be lost, if any, during the propagation of the optical pulse. In that case, CSRM would be used to find the self-localized states of the system with the same accuracy of the SRM. Also one can intentionally undersample the sparse optical signal with monochromators are only used for selected components, but the accurate self-localized state can still be reconstructed using CSRM using only those selected components. It is possible to make the selection of $M$ components deterministically as well, but in that case CS solution would produce some replicated patterns in the solution which need to be filtered in that case \cite{Baysci}.

A similar procedure based on CS which exploits the sparsity of the simulated signal is proposed for general spectral schemes in \cite{Baysci, Baytrbz2, BayTWMS2015, Baytrbz1}. In these works, the main advantage of using CS in a spectral scheme is to improve the numerical simulation time and computer memory requirements of the sparse signals. CSRM can also be computationally advantageous depending on how large $N-M$ is and the number of iteration steps needed to obtain a convergent solution but for a more general SRM this advantage would be more clear. If the ansatz $\zeta(x,z)=\eta(x,\mu) \textnormal{exp}(i\mu(z) z)$, where $\mu(z)$ shows a propagation distance dependent soliton eigenvalue, would be used, then finding the self-localized state at any $z$ would be necessary and thus CSRM would also be computationally very advantageous while there is negligible accuracy difference between CSRM and SRM.

Although sparsity property of the self-localized states of the model NLS-like equation in the physical domain is used and random selection of $M$ components are done in Fourier space, for a state which is sparse in Fourier domain random samples can be taken in the physical domain and the CSRM can be applied in a pseudospectral manner. Additionally the CSRM method can be extended to other spectral methods, such as wavelets, DCTs, Chebyshev and Legendre polynomials, just to name a few.

\section{\label{sec:level1}Results and Discussion}
 
\subsection{Single and Dual Soliton Solutions of the NLS Equation for Zero Optical Potential}
%\begin{doublespace}
It is well known that NLS equation, which can be obtained by setting $V=0$ and $N(\left| \zeta \right|^2)=\left| \zeta \right|^2$ in Eq.~(\ref{eq01}), admits single and dual humped soliton solutions in the form of sech functions. In this section we assess the accuracy and advantages of the proposed CSRM using these single and dual soliton solutions. The parameters of the computations are selected as $p=10, \mu=0.8$. 

\begin{figure}[htb!]
\begin{center}
   \includegraphics[width=3.4in]{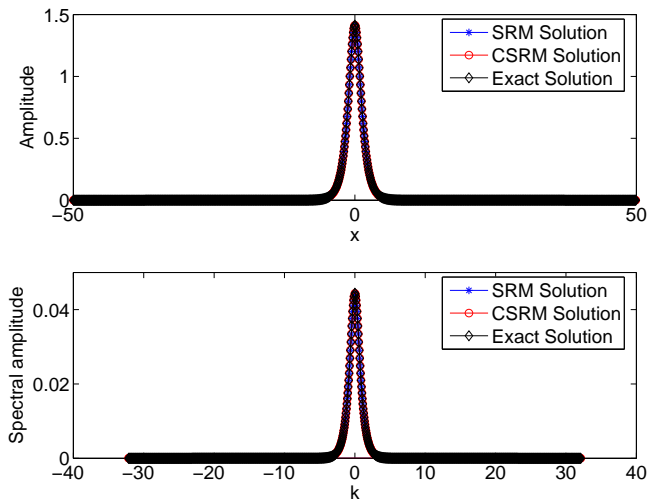}
  \end{center}
\caption{\small Comparison of exact,  $N=1024$ component SRM and $M=256$ component CSRM solutions for a) single soliton b) their spectra }
  \label{fig1}
\end{figure}

In Figure~\ref{fig1}, the $N=1024$ component SRM and $M=256$ component CSRM are compared with the exact single soliton solution of the NLS equation. The initial condition for this simulation is simply a Gaussian in the form of $\exp{(-x^2)}$. The convergence is defined as the normalized change of $\alpha$ is less than $10^{-15}$. Both the SRM and CSRM converge to the exact single sech type soliton solution within few iteration steps. Both of the SRM and CSRM are in excellent agreement with the exact solution as it can be seen in the figure. The normalized root-mean-square error calculated using the exact single sech type solution and CSRM solution is $7.74x10^{-5}$ in the physical domain and is $7.70x10^{-5}$ in the Fourier domain. The two methods are in excellent agreement as it can be seen in Figure~\ref{fig1} while CSRM is more advantageous against missing spectral data since it uses only $M=256$ components.

\begin{figure}[htb!]
\begin{center}
   \includegraphics[width=3.4in]{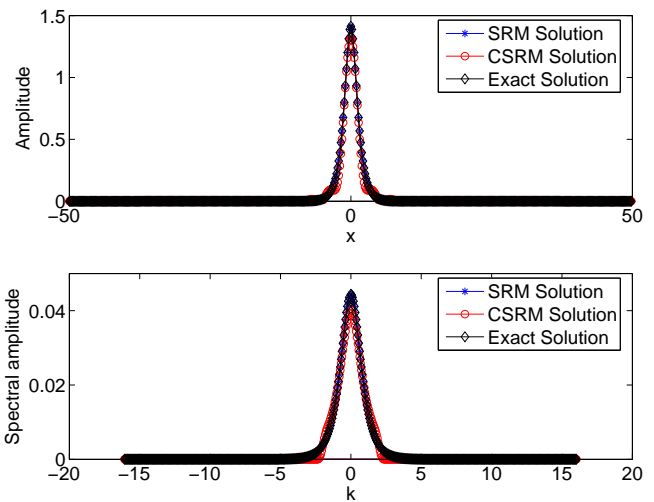}
  \end{center}
\caption{\small Comparison of exact,  $N=512$ component SRM and $M=64$ component CSRM solutions for a) single soliton b) their spectra }
  \label{fig2}
\end{figure}

In Figure~\ref{fig2}, the $N=512$ component SRM and $M=64$ component CSRM are compared with the exact single soliton solution of the NLS equation. The initial condition for this simulation is again a Gaussian. As before, the convergence is defined as the normalized change of $\alpha$ to be less than $10^{-15}$ and both the SRM and CSRM converge to the exact single sech type soliton solution within few iteration steps. Both of the SRM and CSRM are in good agreement with the exact solution as shown in figure. The normalized root-mean-square error for this case, which is calculated using the exact single sech type solution and CSRM solution, is $1.42x10^{-2}$ in the physical domain and is $2.00x10^{-2}$ in the Fourier domain. Compared to the results of the previous case depicted in Figure~\ref{fig1}, error slightly increases due to increased undersampling ratio ($N/M$). The two methods are in good agreement as it can be seen in Figure~\ref{fig2} while CSRM is again more advantageous and robust against missing spectral data than the SRM since it uses only $M=64$ components for the reconstruction of the self-localized state in the form of single soliton.

\begin{figure}[htb!]
\begin{center}
   \includegraphics[width=3.4in]{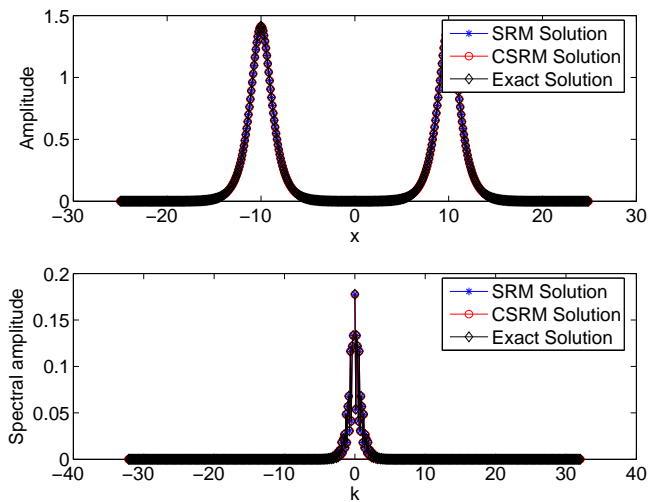}
  \end{center}
\caption{\small Comparison of exact,  $N=512$ component SRM and $M=128$ component CSRM solutions for a) dual soliton b) their spectra }
  \label{fig3}
\end{figure}

$N=512$ component SRM and $M=128$ component CSRM are compared with the dual soliton solution of the NLS in Figure~\ref{fig3}. The initial condition for this simulation is simply superposition of two Gaussians in the form of $\exp{(-(x-x_0)^2)}+\exp{(-(x-x_1)^2)}$ where $-x_0=x_1=10$. The convergence for dual soliton simulations is defined as the normalized change of $\alpha$ to be less than $10^{-5}$, since a smaller error bound may lead to single soliton solution. Both the SRM and CSRM converges to the exact dual sech type soliton solution within few iteration steps. Both of the methods are in excellent agreement as it can be seen in the figure. The normalized root-mean-square error calculated using the exact single sech type solution and CSRM solution is $1.55x10^{-4}$ in physical domain  and $5.45x10^{-5}$ in the Fourier domain. Compared to the case with same $N$ and $M$ depicted in Figure~\ref{fig1}, the slight increase in the error is due to dual soliton profile is wider than single soliton profile, thus has less zero entries which starts to violate the sparsity condition of the CS.

\begin{figure}[htb!]
\begin{center}
   \includegraphics[width=3.4in]{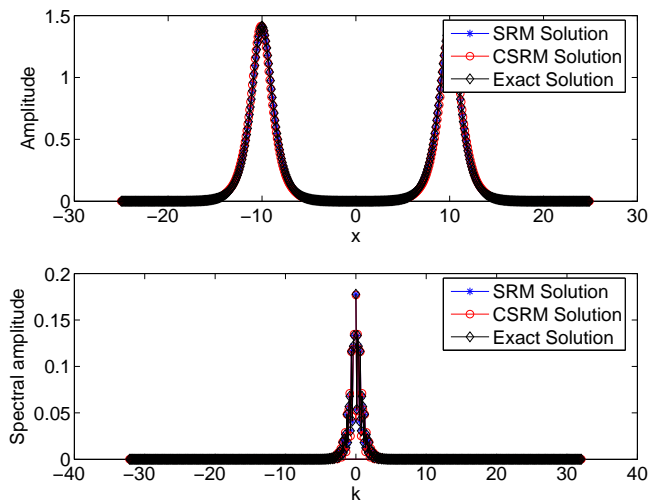}
  \end{center}
\caption{\small Comparison of exact,  $N=512$ component SRM and $M=64$ component CSRM solutions for a) dual solitons b) their spectra }
  \label{fig4}
\end{figure}

We compare the $N=512$ component SRM and the $M=64$ component CSRM with the exact dual soliton solution of the NLS in Figure~\ref{fig4}. The initial condition for this simulation is again superposition of two Gaussians with unit amplitudes located at $-10$ and $10$. As before, the upper bound for the convergence criteria of $\alpha$ is selected as $10^{-5}$. Both the SRM and CSRM converges to the exact dual sech type soliton solution within few iteration steps. For this case, the normalized root-mean-square error calculated using the exact dual sech type soliton solution and CSRM solution is $1.32x10^{-2}$ in physical domain and $4.60x10^{-3}$ in the Fourier domain. The two methods are in acceptable agreement as it can be seen in the figure. Compared to the case with same $N$ and $M$ depicted in Figure~\ref{fig3}, the increase in the error is due to the dual soliton profile that has less zero values compared to single soliton, which causes the sparsity condition of the CS to be violated. It is natural to expect that with higher undersampling ratios (i.e. more missing spectral data), the error will increase and CSRM may eventually fail. However the capability of CSRM to capture self-localized solutions despite these large undersampling ratios shows that it can be a very useful method in evaluating self-localized solutions in many systems with missing spectral data.

\subsection{Single and Dual Soliton Solutions of NLS-like Equation for Nonzero Optical Potential}
Turning our attention to the NLS-like equation given in Eq.~(\ref{eq01}) for a nonzero optical potential of $V=I_o \cos^2(x)$ and the nonlinear term given as $N(\left| \eta \right|^2)=-1/(1+\left| \eta \right|^2)$ which are used in practice \cite{Segev}, the iteration formula becomes Eq.~(\ref{eq09}). In this iteration formula the parameters are selected as $I_0=1, p=10, \mu=0.8$. 

\begin{figure}[htb!]
\begin{center}
   \includegraphics[width=3.4in]{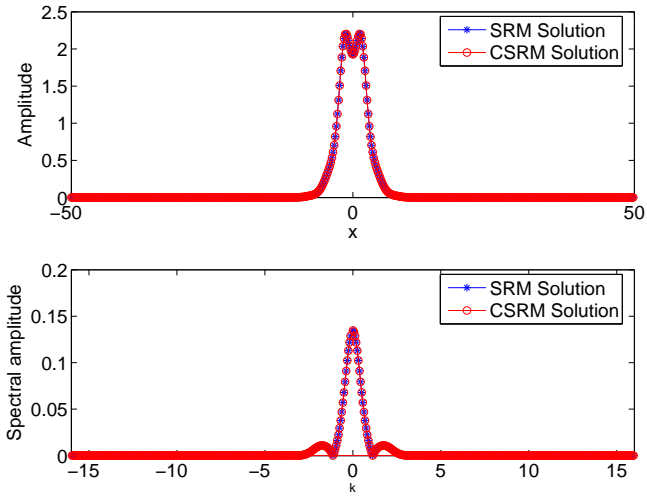}
  \end{center}
\caption{\small Comparison of $N=512$ component SRM and $M=128$ component CSRM solutions for a) single soliton b) their spectra }
  \label{fig5}
\end{figure}

In Figure~\ref{fig5}, $N=512$ component SRM and $M=128$ component CSRM solutions are compared with each other. The initial condition for this simulation is simply a Gaussian in the form of $\exp{(-x^2)}$. The convergence is defined as the normalized change of $\alpha$to be less than $10^{-15}$. Both the SRM and CSRM converge to the solution shown in Figure~\ref{fig5} within few iteration steps. The SRM and CSRM solutions are in excellent agreement as one can see in the figure. The normalized root-mean-square difference calculated using the SRM solution and CSRM solution is $2.47x10^{-4}$ in the physical domain and is $1.14x10^{-4}$ in the Fourier domain. The accuracy difference is of negligible importance but the CSRM is more advantageous and robust against missing spectral data since it uses only $M=128$ components for the reconstruction of the single soliton solution.

\begin{figure}[htb!]
\begin{center}
   \includegraphics[width=3.4in]{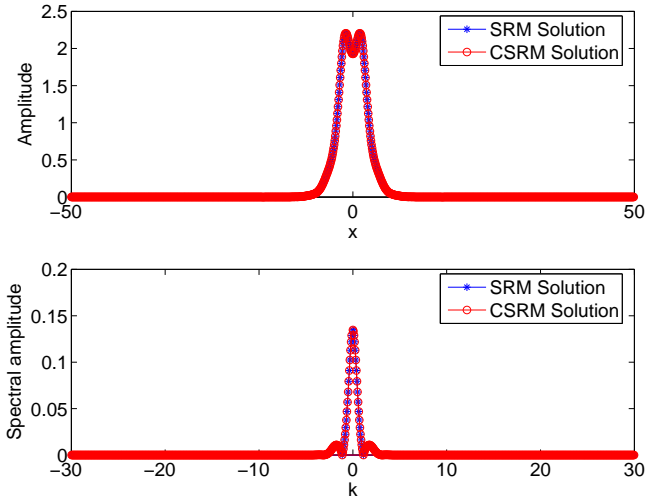}
  \end{center}
\caption{\small Comparison of $N=1024$ component SRM and $M=128$ component CSRM solutions for a) single soliton b) their spectra }
  \label{fig6}
\end{figure}

In Figure~\ref{fig6}, $N=1024$ component SRM and $M=128$ component CSRM solutions are compared with each other. The initial condition for this simulation is simply a Gaussian. The convergence is defined as the normalized change of $\alpha$ to be less than $10^{-15}$. Both the SRM and CSRM converge to the solution shown in Figure~\ref{fig6} within few iteration steps. The SRM and CSRM solutions are in excellent agreement as one can see in the figure. The normalized root-mean-square difference calculated using the SRM solution and CSRM solution is $2.48x10^{-4}$ in the physical domain and is $8.12x10^{-5}$ in the Fourier domain. The accuracy difference is of negligible importance but the CSRM is more advantageous and robust against missing spectral data since it uses only $M=128$ components for the reconstruction of the single soliton solution.

\begin{figure}[htb!]
\begin{center}
   \includegraphics[width=3.4in]{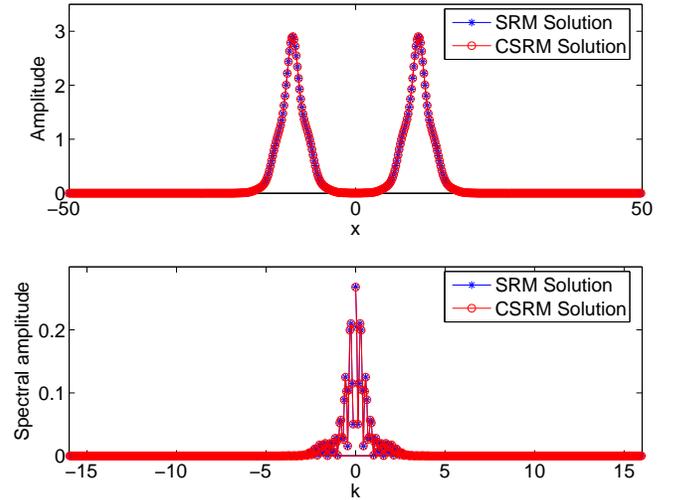}
  \end{center}
\caption{\small Comparison of $N=512$ component SRM and $M=128$ component CSRM solutions for a) dual soliton b) their spectra }
  \label{fig7}
\end{figure}

In Figure~\ref{fig7}, $N=512$ component SRM and $M=128$ component CSRM solutions are compared with each other. The initial condition for this simulation is simply superposition of two Gaussian in the form of $\exp{(-(x-x_0)^2)}+\exp{(-(x-x_1)^2)}$ where $-x_0=x_1=10$. The convergence for criteria for dual soliton simulations is selected as the normalized change of $\alpha$ to be less than $10^{-5}$, since a smaller error bound may lead to single soliton solution. Both the SRM and CSRM converge to the solution shown in Figure~\ref{fig7} after few iterations. The SRM and CSRM solutions are in good agreement as depicted in the figure. The normalized root-mean-square difference calculated using the SRM solution and CSRM solution is $4.27x10^{-4}$ in the physical domain and is $9.91x10^{-5}$ in the Fourier domain. The accuracy difference is of negligible importance but the CSRM is more advantageous and robust against missing spectral data since it uses only $M=128$ components for the reconstruction of the single soliton solution.

\begin{figure}[htb!]
\begin{center}
   \includegraphics[width=3.4in]{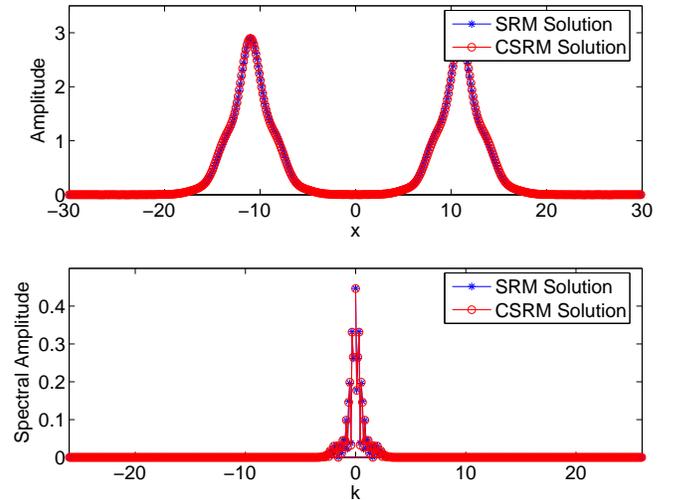}
  \end{center}
\caption{\small Comparison of $N=512$ component SRM and $M=64$ component CSRM solutions for a) dual soliton b) their spectra }
  \label{fig8}
\end{figure}

For the last assessment, we compare the $N=512$ component SRM and the $M=64$ component CSRM solutions of the NLS-like equation with a self focusing saturable nonlinearity in Figure~\ref{fig8}. The initial condition for this simulation is again superpositions two Gaussian with unit amplitudes located at $-10$ and $10$. As before, the upper bound for the convergence criteria of $\alpha$ is selected as $10^{-5}$. Both the SRM and CSRM converges to the depicted solution after few iteration steps. For this case, the normalized root-mean-square difference using the SRM and the CSRM solutions is $4.80x10^{-3}$ in physical domain and $1.40x10^{-3}$ in the Fourier domain. The two methods are in acceptable agreement as it can be seen in the figure. Compared to the previous cases, the increase in the error is due to the less number of zeros in the wider dual soliton profile compared to single soliton, which causes the sparsity condition of the CS to be violated. The potential term has also an effect on the violation of CS since it affects the soltion shapes. It is natural to expect that with higher undersampling ratios (i.e. more missing spectral data), the error will increase and CSRM may eventually fail as discussed before for the case depicted in Figure~\ref{fig4}. However the capability of CSRM to capture self-localized solutions despite these large undersampling ratios shows that CSRM can be a very useful method in evaluating self-localized solutions in many systems with missing spectral data. This method can also be generalized to many other nonlinear system of equations used to describe many different physical phenomena and can also be applied to other periodic or nonperiodic domain spectral methods for evaluating computational solutions under missing spectral or pseudospectral data.

% - figures and tables

\section{\label{sec:level1}Conclusion and Future Work}

In this paper we have proposed a novel numerical scheme for finding the sparse self-localized states of a nonlinear system of equations when there is missing spectral data. The method utilizes far fewer number of spectral components ($M$) than classical versions of the Petviashivili's method or spectral renormalization method with higher number of spectral components ($N$). After the converge criteria is achieved for $M$ components, the signal with $N$ components can be reconstructed from $M$ components by the $l_1$ minimization technique of the compressive sampling. This method can be named as the compressive spectral renormalization (CSRM) method. Compared to SRM, the main advantage of the CSRM is that, it is capable of finding the sparse self-localized states of the evolution equation(s) with far fewer spectral data. For example for fiber optical communications, where some data is lost during the propagation of the optical pulse or considering memory and time constraints they may be intentionally ignored, CSRM can be used to find self localized states of the system of equations studied. CSRM can also be computationally advantageous depending on how large $N-M$ is and the number of iteration steps needed to obtain a convergent solution but for a more general extension of SRM, where soliton eigenvalue depends of propagation distance, this advantage would be more clear. For such a case application of SRM at each along fiber point would be necessary thus CSRM would also be computationally very advantageous while there would be negligible accuracy difference between CSRM and SRM.

There are some sparse FFT algorithms well developed and used in the literature. As a future work it is possible to implement these sparse fast transforms for computational modeling of the sparse signals and provide a comparison with the CSRM. The sequential, parallel or distributed algorithms can be used for this purpose. The CSRM can also be incorporated for other type of spectral methods such as those where the wavelets, DCTs, Legendre, Chebyshev and other forms of basis functions are used for computational simulations.

\end{document}